\documentclass[
     ,final            
  ]
  {aipproc}

\layoutstyle{6x9}
\usepackage{amssymb}

\begin{document}

\title[3D MHD Simulations of accreting NSs: evidence of QPO emission from the surface.]{3D MHD Simulations of accreting neutron stars: evidence of QPO emission from the surface.}


\author{Matteo Bachetti}{
  address={Universit\`a degli Studi di Cagliari}
}

\author{Marina M. Romanova}{
  address={Cornell University}
}

\author{Akshay Kulkarni}{
  address={Cornell University}
}

\author{Luciano Burderi}{
  address={Universit\`a degli Studi di Cagliari}
}

\author{Tiziana di Salvo}{
  address={Universit\`a degli Studi di Palermo}
}

\begin{abstract}
3D Magnetohydrodynamic simulations show that when matter accretes onto neutron stars, in particular if the misalignment angle is small, it does not constantly fall at a fixed spot. Instead, the location at which matter reaches the star moves. These moving hot spots can be produced both during stable accretion, where matter falls near the magnetic poles of the star, and unstable accretion, characterized by the presence of several tongues of matter which fall on the star near the equator, due to Rayleigh-Taylor instabilities. Precise modeling with Monte Carlo simulations shows that those movements could be observed as high frequency Quasi Periodic Oscillations. We performed a number of new simulation runs with a much wider set of parameters, focusing on neutron stars with a small misalignment angle. In most cases we observe oscillations whose frequency is correlated with the mass accretion rate $\dot{M}$. Moreover, in some cases double QPOs appear, each of them showing the same correlation with $\dot{M}$. 
\end{abstract}

\maketitle

\begin{center}
 {\bf \uppercase{ Moving hot spots and QPOs}}
\end{center}

In the last few years 3D MHD simulations have shown that in many cases, in particular if the misalignment angle is small ($\lesssim 15^{\circ}$), matter falls in a ring around the magnetic pole. The brightest spot (the hot spot) of the ring is not fixed, but moves with an angular velocity related to the Keplerian velocity  in a zone of the disk outside the magnetospheric radius. This is observed both in the case of stable accretion (e.g. \cite{Romanova:2004}) and in the case of unstable accretion (e.g. \cite{Kulkarni:2008},\cite{Romanova:2008}).
During unstable accretion the hot spots produced by instabilities in the equatorial region move, with a velocity similar to that of the inner disk (\cite{Bachetti:2009},\cite{Romanova:2009}). The motion in this case is less regular and shorter. We studied the effects of these movements in detail with Monte Carlo techniques and showed that the features produced by those movements are quasi periodic oscillations.

\smallskip
\begin{center}
{\bf Correlations with the mass accretion rate}
\end{center}
\begin{figure}
\centering
$ \begin{array}{cc}
 \includegraphics[width=2.5in]{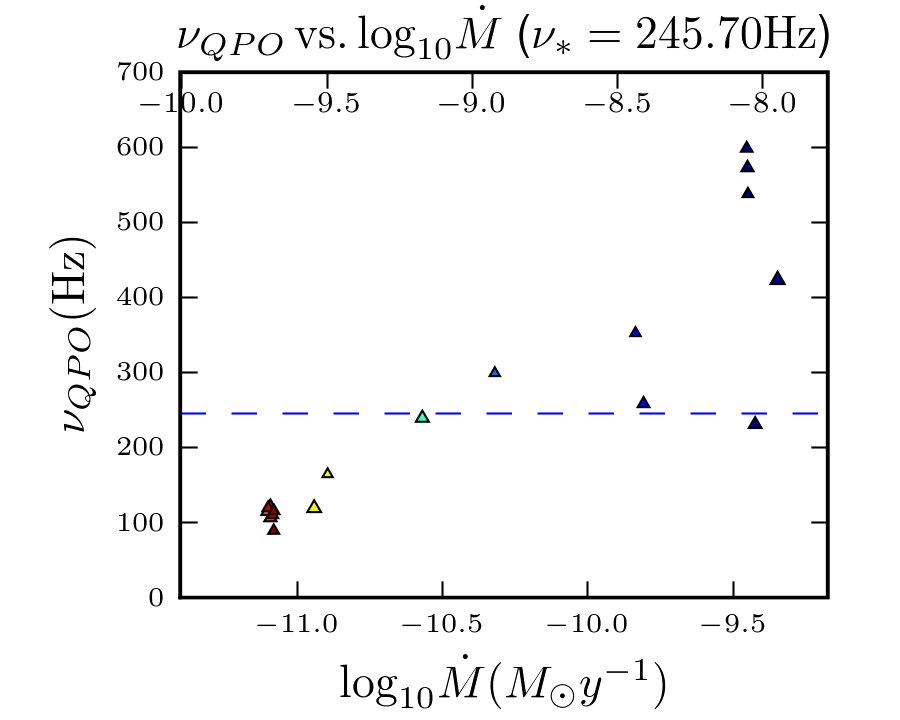} &
 \includegraphics[width=2.5in]{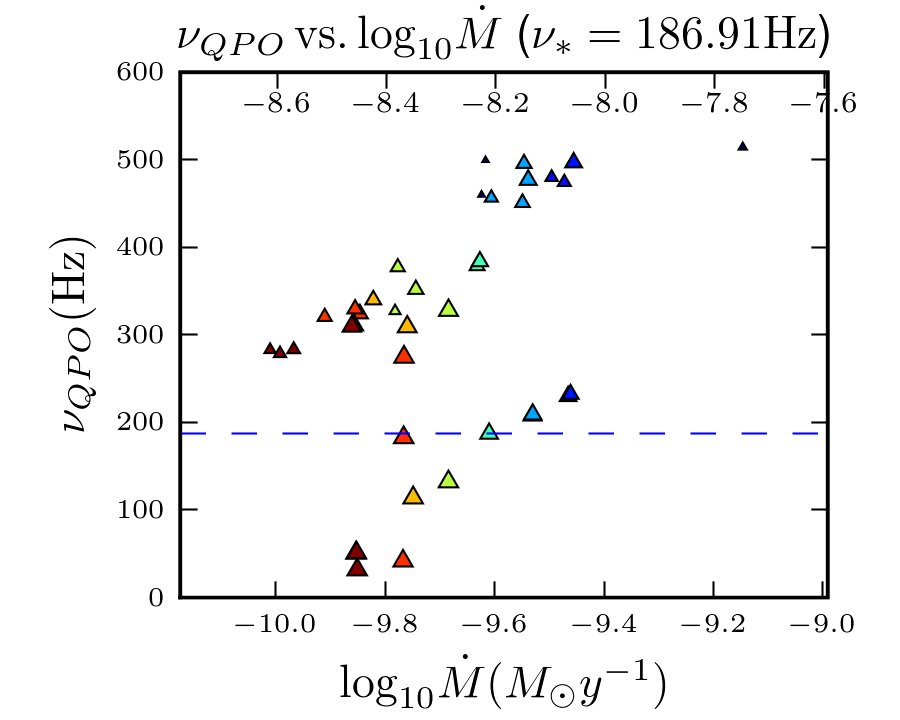}
 \end{array} $
\caption{Frequency variation of the hot spot movements with the mass accretion rate \cite{Bachetti:2009}. Results are scaled for the two cases $B=10^8$G and $B=5\cdot10^8$G. The first case (right) shows stable accretion from the magnetic pole, with matter falling onto the surface along the funnel formed by the magnetic field lines. The hot spot moves  around the magnetic pole with a certain angular velocity, which we plot against the mass accretion rate $\dot{M}$, finding a good correlation.
The second case (left) presents instabilities at the magnetospheric boundary (or truncation radius). In this case we have both accretion onto the magnetic pole with the usual funnel-like stream, that moves as in the first case, and accretion from tongues that pierce the magnetosphere falling near the equator, moving approximately at the frequency of the inner disk. Hence the ``double track'' in the plot.}
\label{fig:numdot}
\end{figure}

We performed a number of simulations in which the only parameter changed was the mass accretion rate.
All the systems reproduced in the simulations have $M=1.4M_{\odot}$, $R=10$km and misaligment angle $\theta=2^{\circ}$. One system has a rotational period $\tau=4.1$ms, the other $\tau=5.4$ms. In Fig.~\ref{fig:numdot} we see the rotational frequency of the hot spots plotted against the mass accretion rate $\dot{M}$. In one case the accretion is only on the poles, in the other both in the poles and in the equatorial zones due to instabilities. In the latter case, the hot spots appearing in the two accretion zones move together with the accretion rate.
This result is, of course, extremely interesting to interpret the presence of multiple QPOs in accreting neutron stars. We have a mechanism to produce very bright {\em pairs} of QPOs, whose spectrum can easily be different from the background emission from the disk, and whose characteristic frequencies change together with $\dot{M}$. Moreover, the quality factors of the two peaks show a different behavior, as observed in many works (e.g. \cite{diSalvo:2003},\cite{Barret:2006}). Submitted to MNRAS. A preprint is available \cite{Bachetti:2009}.


\begin{theacknowledgments}
Research supported by INFN (MB) and the contract ASI-INAF I/088/06/0 for High Energy Astrophysics (MB, LB and TD).
MMR and AK were supported in part by NASA grant NNX08AH25G and by NSF grant AST-0807129. Simulations performed on supercomputers by NASA (Pleiades, Columbia), and by CINECA (BCX, under the 2008-2010 INAF/CINECA agreement). 
\end{theacknowledgments}



\bibliographystyle{aipproc}   


\IfFileExists{\jobname.bbl}{}
 {\typeout{}
  \typeout{******************************************}
  \typeout{** Please run "bibtex \jobname" to optain}
  \typeout{** the bibliography and then re-run LaTeX}
  \typeout{** twice to fix the references!}
  \typeout{******************************************}
  \typeout{}
 }

\end{document}